%% file: ms.tex
\documentclass[conference]{IEEEtran}
\IEEEoverridecommandlockouts
\usepackage{cite}
\usepackage{amsmath,amssymb,amsfonts}
\usepackage{algorithmic}
\usepackage{graphicx}
\usepackage{textcomp}
\usepackage[table]{xcolor}

\def\BibTeX{{\rm B\kern-.05em{\sc i\kern-.025em b}\kern-.08em
    T\kern-.1667em\lower.7ex\hbox{E}\kern-.125emX}}
\usepackage[symbol]{footmisc}

\definecolor{tablegray}{gray}{0.9}

\usepackage{subcaption}
\usepackage{tikz}
\usepackage{placeins}
\usepackage{wrapfig}
\usepackage{pgfplots}
\usepackage{boldline}
\pgfplotsset{compat=1.5}
\usepackage{booktabs}
\usepackage{makecell}
\usepackage{multirow}
\usepackage{verbatim}
\usepackage{placeins}
\usepackage{bbm}
\usepackage[linesnumbered,lined,commentsnumbered, ruled]{algorithm2e}
\usepackage{hyperref}

\begin{document}

\title{
Stay Ahead of Poachers: Illegal Wildlife Poaching Prediction and Patrol Planning Under Uncertainty with Field Test Evaluations
\thanks{*Equal contribution.}
}


\author{
\IEEEauthorblockN{
Lily~Xu*{$^1$},
Shahrzad~Gholami*{$^2$},
Sara~Mc~Carthy{$^2$},
Bistra~Dilkina{$^2$},
Andrew~Plumptre{$^3$},
Milind~Tambe{$^1$}, \\
Rohit Singh{$^4$},
Mustapha Nsubuga{$^5$},
Joshua Mabonga{$^5$}, \\
Margaret~Driciru{$^6$},
Fred~Wanyama{$^6$},
Aggrey~Rwetsiba{$^6$},
Tom~Okello{$^6$},
Eric~Enyel{$^6$}
}
\IEEEauthorblockA{
\textit{
$^1$Harvard University, 
$^2$University of Southern California, 
$^3$Key Biodiversity Area Secretariat,} \\ 
\textit{
$^4$World Wide Fund for Nature, 
$^5$Wildlife Conservation Society, 
$^6$Uganda Wildlife Authority
} \\
\texttt{
lily\_xu@g.harvard.edu,
sgholami@usc.edu} 
}
}

\maketitle

\input{Abstract}

\begin{IEEEkeywords}
wildlife protection, data mining, predictive modeling, patrol route planning, poaching, uncertainty, gaussian processes
\end{IEEEkeywords}

\input{1-Introduction}

\input{2-RelatedWork}

\input{3-Domain}

\input{4-PredictiveModeling}

\input{5-EvaluatePredictiveModeling}
\input{7-PatrolPlanning}
\input{6-FieldTests}
\input{8-Conclusion}

\FloatBarrier
\bibliographystyle{IEEEtran}
\bibliography{refs.bib}

\end{document}

%% file: Abstract.tex
\begin{abstract}
Illegal wildlife poaching threatens ecosystems and drives endangered species toward extinction. However, efforts for wildlife protection are constrained by the limited resources of law enforcement agencies.
To help combat poaching, the Protection Assistant for Wildlife Security (PAWS) is a machine learning pipeline that has been developed as a data-driven approach to identify areas at high risk of poaching throughout protected areas and compute optimal patrol routes. In this paper, we take an end-to-end approach to the data-to-deployment pipeline for anti-poaching. In doing so, we address challenges including extreme class imbalance (up to 1:200), bias, and uncertainty in wildlife poaching data to enhance PAWS, and we apply our methodology to three national parks with diverse characteristics. 
(i)~We use Gaussian processes to quantify predictive uncertainty, which we exploit to improve robustness of our prescribed patrols and increase detection of snares by an average of 30\%. We evaluate our approach on real-world historical poaching data from Murchison Falls and Queen Elizabeth National Parks in Uganda and, for the first time, Srepok Wildlife Sanctuary in Cambodia.
(ii)~We present the results of large-scale field tests conducted in Murchison Falls and Srepok Wildlife Sanctuary which confirm that the predictive power of PAWS extends promisingly to multiple parks. This paper is part of an effort to expand PAWS to 800~parks around the world through integration with SMART conservation software. 

\end{abstract}





%% file: 1-Introduction.tex
\section{Introduction}

Illegal wildlife poaching is an international problem that threatens biodiversity, ecological balance, and ecotourism~\cite{cooney2017poachers}. Countless species are being poached to near-extinction: the ivory, horn, and skin of exotic species such as elephants, rhinos, and tigers render them targets for illegal trade of luxury products and medicinal applications~\cite{chase2016continent,spillane2015africa}; other animals like wild pigs and apes are hunted as bushmeat for protein~\cite{warchol2004transnational}. Even when their habitats become designated wildlife conservation areas, these animals continue to be at risk due to lack of sufficient resources to protect them from poachers. Timely detection and deterrence of illegal poaching activities in protected areas are critical to combating illegal poaching. Artificial intelligence frameworks can significantly advance wildlife protection efforts by learning from past poaching activity to prescribe actionable recommendations to park managers. 

\begin{figure}[t!]
\centering
\includegraphics[width=.9\columnwidth]{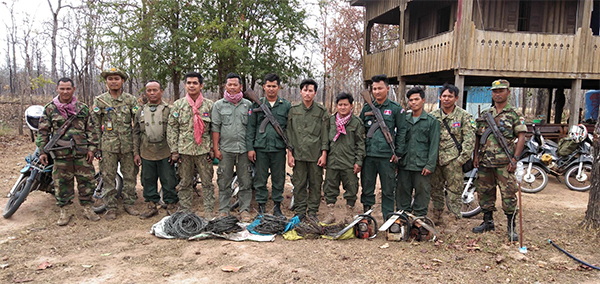}
\caption{Rangers in Srepok Wildlife Sanctuary with snares they removed during our field tests in December 2018. Photo: WWF Cambodia.}
\label{fig:snares-cambodia}
\end{figure}

\begin{figure}
\centering
\includegraphics[width=.9\columnwidth]{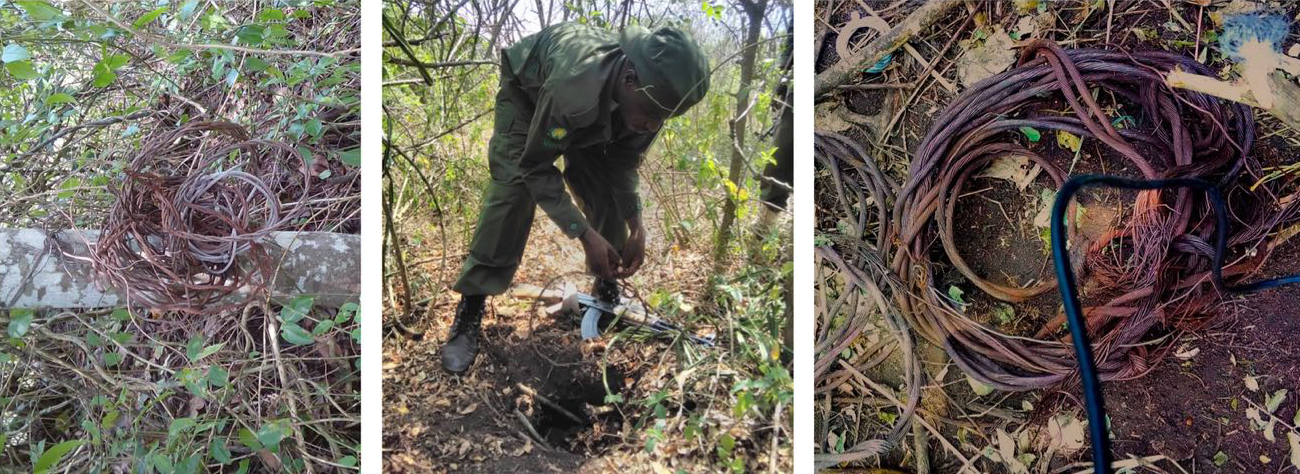}
\caption{Well-hidden snares detected and removed by rangers during our field tests in Murchison Falls National Park in Uganda. Photos: Uganda Wildlife Authority.}
\label{fig:traps}
\end{figure}

Assessing poaching risk through a protected area and prescribing patrol plans to rangers requires in-depth knowledge of the poachers' behavior. Learning the poachers' behavior is a challenging machine learning problem since 
(a)~the wildlife crime datasets are typically extremely imbalanced, with up to 99.6\% negative labels;
(b)~negative labels indicating absence of illegal activity are not reliable due to the inherent difficulty of detecting well-hidden poaching signs in the forest, as shown in Fig.~\ref{fig:traps};
(c)~historical poaching observations are not collected thoroughly and uniformly, resulting in biased datasets; and
(d)~poaching patterns and landscape features vary from one protected area to another, so a universal predictive model cannot be recommended. These challenges with unreliable and imbalanced data are pervasive in machine learning, thus the methodology proposed in this paper can be extended to other domains with real-world data. 


Traditionally, the data-to-deployment pipeline in AI begins by feeding data into a predictive machine learning algorithm to generate predictions; those predictions feed into a prescriptive component that computes decisions based on the predictions; and finally those decisions are deployed. This traditional approach has been used in past work for wildlife conservation: the Protection Assistant for Wildlife Security (PAWS), introduced in \cite{yang2014adaptive}, is a two-stage approach for anti-poaching. The first stage employs machine learning to build a predictive model of relative poaching risk, and the second stage leverages these predictions in a game theoretic model to optimize the detection of future poaching activity and prescribe patrol routes. Previous work considers each of these components (data, prediction, prescription, and deployment) as separate. However, the four wildlife domain challenges enumerated above demand an end-to-end approach to the data-to-deployment pipeline. 

In this paper for the first time, we show a closer integration of the different components in the pipeline\footnote[2]{Our code is available online at \url{https://github.com/lily-x/paws-public}}. Specifically, the entire pipeline of this paper has been designed with deployment in mind, striving to offer risk-averse patrols to rangers. In the anti-poaching domain, the ultimate goal is to prescribe effective patrol routes for rangers to maximize the number of snares removed---corresponding to animal lives saved. Thus, rangers are incentivized to conduct patrols with higher certainty of detecting snares. This domain insight inspired us to optimize the patrol plans by measuring the predictive uncertainty in our model. To do so, (i)~We use Gaussian processes to quantify uncertainty in predictions of poaching risk and exploit these uncertainty metrics in our optimization problem to increase the robustness of our prescribed patrols. We evaluate our approaches on historical crime data using three real-world datasets from Uganda and Cambodia, which have different characteristics from both ecological and data quality perspectives.
(ii)~We present results from large-scale field tests conducted in Murchison Falls National Park in Uganda and Srepok Wildlife Sanctuary in Cambodia, which demonstrate that the power of our model to predict risk of future poaching activity extends promisingly to the real world.  


To deploy these anti-poaching models, we have worked with rangers and conservation specialists at the World Wild Fund for Nature~(WWF), Wildlife Conservation Society~(WCS), and Uganda Wildlife Authority~(UWA). During our field tests, rangers detected 38~attacked cells in MFNP and removed over 1,000 snares from SWS in a single month. Along with evaluating the predictive power of our model, these tests enable us to garner feedback from rangers and convince these conservation specialists that a machine learning approach can augment the wildlife protection strategies they employ.

The historical patrolling and poaching data for the three parks were collected over the years using SMART, a monitoring and reporting tool for protected area management~\cite{smart}. SMART is developed by a consortium of leading conservation organizations and used in more than 800 protected areas across 55 countries. Although SMART records significant amounts of historical data, its current capabilities are limited to managing data; the missing link is to leverage that data to inform patrol strategy. To address this limitation, PAWS will be integrated into the SMART software in the coming year and become available to park managers around the world. The enhancements and field tests in this paper outline significant steps to expand PAWS on a global scale.

%% file: 2-RelatedWork.tex
\section{Related Work} 




In the anti-poaching predictive modeling literature, \cite{yang2014adaptive}~introduced PAWS as a stochastic behavioral model; the project has since grown into a multi-year, multi-university effort. Previous work in PAWS can be divided into two main threads, conceptual and practical; we also discuss relevant work in PU learning and machine learning under uncertainty. 

\paragraph{Conceptual thread of PAWS}
The origins of PAWS were developed in conceptual papers that provided the theoretical foundations using game theory and machine learning. CAPTURE~\cite{nguyen2016capture} used a two-layered Bayesian network with latent variables to model imperfect detection of poaching activity. In contrast, INTERCEPT~\cite{kar2017cloudy} was proposed as an ensemble of decision trees that did not assume imperfect detection of poaching activities but achieved better runtime and performance than CAPTURE. Combining the strengths of the previous approaches, \cite{gholami2017taking} introduced a geo-clustering technique to produce a hybrid model consisting of multiple Markov random fields and a bagging ensemble of decision trees. The patrol planning component of PAWS is developed through a game theoretical framework~\cite{fang2015introduction}, which extends the game theoretic models of Stackelberg Security Games \cite{tambe2011security, korzhyk2011solving} to the setting of environmental sustainability and wildlife protection, known as Green Security Games (GSGs). In this setting, adversaries may not behave as perfectly rational utility maximizers. GSGs have been successfully applied domains such as to logging~\cite{johnson2012patrol} and fishing~\cite{Haskell_IAAI2014}. 

This conceptual development continues today, with online approaches attempting to close the loop between data gathering and patrol planning. \cite{gholami2019strategies} attempts to address the bias in historical data, proposing an online algorithm that balances a patrol-planning model trained with historical data against a model with no prior knowledge to determine the usefulness historical data, and \cite{wang2019deep} uses deep reinforcement learning to respond to real-time information such as footprints. However, these conceptual models are unable to feed back into the real-world domain that inspired the work: they do not use any real-world data, and the models are computationally intractable in practice in large parks with thousands of targets. 

\paragraph{Practical thread of PAWS}
To conduct the first field test of predictive modeling for poaching, \cite{kar2017cloudy} worked with park rangers in Queen Elizabeth National Park to suggest target regions for one month of patrol. However, the results of that trial are inconclusive, as there was no control group. \cite{gholami2017taking} extended field tests in QENP, which more convincingly demonstrated the predictive power of the model, but these tests were limited: they had only two experiment groups (high and low risk), and only 9\% of the areas used in the field test were assessed to be high-risk of poaching. In this paper, we present more extensive field tests in two additional national parks with three risk categories (high, medium, and low) instead of two. In addition, neither of \cite{kar2017cloudy, gholami2017taking} designed the work with deployment in mind. By contrast, we developed the methodology of this paper with the intent to aid deployment.

The iWare-E ensemble model~\cite{gholami2018adversary} addresses one major challenge of anti-poaching predictive modeling: non-uniform, one-sided noise on observations of poaching activity due to imperfect detection of snares and other poaching signs. This ensemble approach exploits other machine learning models as weak learners; in particular, bagging ensembles of decision trees or SVMs were used as weak learners for evaluation. 
However, the model falsely assumes that the attack probabilities as predicted by adversary models are exact and thus can be directly exploited for patrol planning without any robustness concerns. To remedy this gap, we present three enhancements to iWare-E, including explicit quantification of prediction uncertainty. In the patrol planning stage, this enhancement enables taking a risk-aware approach that actively trades off between exploration and exploitation. 

\paragraph{PU learning and uncertainty}
Outside of the conservation domain, research addressing similar data issues have emerged, such as positive-unlabeled (PU) learning and learning under uncertainty. Introduced in \cite{lee2003learning}, PU learning focuses on unreliable negative labels, taking a semi-supervised approach to binary classification. PU learning has been applied to domains such as remote sensing data~\cite{li2010positive} and gene identification~\cite{yang2012positive}. The problem of quantifying uncertainty in machine learning models is often approached by using Bayesian methods, notably Gaussian processes~\cite{rasmussen2004gaussian}, or looking at confidence intervals, such as in random forests~\cite{mentch2016quantifying}. We incorporate Gaussian processes to quantify uncertainty, and compare those results with variance from random forests.

%% file: 3-Domain.tex
\section{Wildlife Crime Domain and Data}
We use wildlife crime datasets from Uganda and Cambodia, provided by UWA, WCS, and WWF. Characteristics of wildlife crime are common in other security domains, such as illegal fishing~\cite{Haskell_IAAI2014} and coastal attacks~\cite{shieh2012protect}, which have also been modeled by Green Security Games. Thus the methodology described in this paper extend to other security domains as well. 

\subsection{Domain Characteristics}

We study Murchison Falls National Park~(MFNP) and Queen Elizabeth National Park~(QENP) in Uganda, and Srepok Wildlife Sanctuary~(SWS) in Cambodia. 
These protected areas cover 5000 sq. km, 2500 sq. km and 4300 sq. km, respectively. MFNP and QENP are critically important for ecotourism and conservation in Uganda, and provide habitat to elephants, giraffes, hippos, and lions~\cite{critchlow2015spatiotemporal}. SWS is the largest protected area in Southeast Asia and is home to elephants, leopards, and banteng. SWS once housed a native population of tigers, but they fell prey to poaching; the last tiger was observed in 2007. In the intervening decade, SWS has been identified as the most promising site in Southeast Asia for tiger reintroduction~\cite{gray2017tigers}. Effectively managing the landscape by reducing poaching will be critical to successful tiger reintroduction.

\begin{figure}
\begin{subfigure}[t]{.31\columnwidth}
\centering
\includegraphics[height=.85\textwidth]{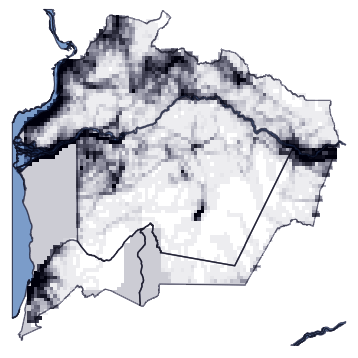}
\caption{MFNP, Uganda}
\label{fig:MFNP}
\end{subfigure}
~
\begin{subfigure}[t]{.31\columnwidth}
\centering
\includegraphics[height=.85\textwidth]{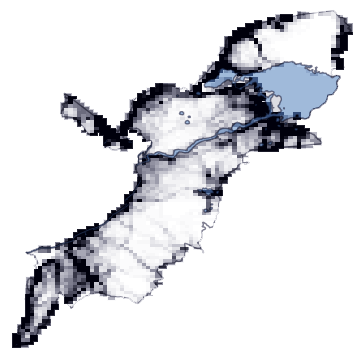}
\caption{QENP, Uganda}
\label{fig:QENP}
\end{subfigure}
~
\begin{subfigure}[t]{.31\columnwidth}
\centering
\includegraphics[height=.85\textwidth]{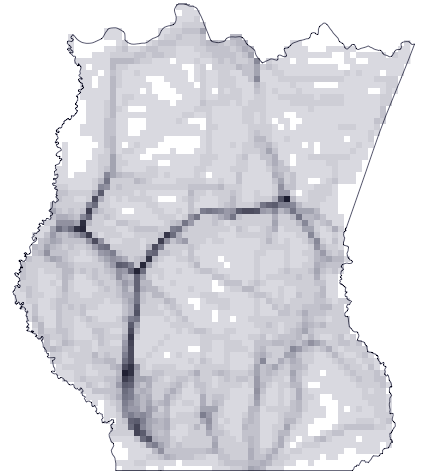}
\label{fig:SWS}
\caption{SWS, Cambodia}
\end{subfigure}
\caption{The protected areas used in this study. Visualized are the historical patrol effort for each protected area, calculated as kilometer patrolled per $1 \times 1$~km cell. Note that patrol effort is unevenly distributed around the park, contributing to bias in the data, and many areas have never been patrolled (in white).}
\label{fig:protectedAreasMap}
\end{figure}

To combat poaching, park rangers conduct patrols through protected areas and use GPS trackers to record their observations. They confiscate animal traps, rescue live animals caught in snares, and arrest any poacher they encounter~\cite{critchlow2016improving}. Their GPS trackers are then synced to the SMART database system~\cite{smart} to manage their many years of wildlife crime data. However, the data are biased due to the inability of rangers to detect all instances of poaching. There are additional data collection issues due to the nature of these patrols: rangers may have to address an emergency in the field, such as hearing a poacher in the distance, and lose the opportunity to record snares or bullet cartridges they found.

\label{sec:cambodia_challenges}
We study SWS in Cambodia for the first time and conduct a four-month field test. This park introduces many challenges. (a)~It is particularly under-resourced, with only a 72~rangers enforcing an area the size of Rhode Island. (b)~Unlike MFNP or QENP where rangers conduct foot patrols, rangers in SWS travel by motorbike. This faster form of transport means that waypoints in the dataset (typically recorded once every 30~minutes) are even more sparse, making it difficult to interpolate their trajectory between sequential points. Additionally, negative labels in the dataset are likely less reliable because patrollers are less able to carefully observe their surroundings when traveling quickly. (c)~SWS experiences strong seasonality: many rivers are impossible to cross during the wet season, but become accessible during the dry season. These distinctions highlight the need to evaluate and improve the PAWS machine learning model by testing in multiple diverse areas, including with field tests to validate the approach on the ground.




\subsection{Dataset Processing}
The features used in our dataset represent static geospatial features about locations within each park. Data specialists at WWF, WCS, and UWA provide us with relevant GIS~shapefiles and GeoTIFF files. The features differ between parks, but include terrain features such as rivers, elevation maps, and forest cover; landscape features such as roads, park boundary, local villages, and patrol posts; and ecological features such as animal density and net primary productivity. We use these static features to build data points in our predictive model, either as direct values (such as slope or animal density) or as distance values (such as distance to nearest river). We do not explicitly encode longitude or latitude as features, which could be extrapolated through the various distance values. 

The dynamic components in the dataset are the historical patrol effort and observed poaching activity. Patrol observations come from SMART conservation software, which records the GPS~location of each observation along with date and time, patrol leader, and method of transport. Rangers enter their observations: animals or humans spotted; signs of illegal activity such as campsites or cut trees; and signs of poaching activity such as firearms, bullet cartridges, snares, or slain animals. We categorize these observations into poaching and non-poaching. Additionally, we rebuild historical patrol effort from these observations by using sequential waypoints to calculate patrol trajectories. 

To study the data, we discretize the protected areas into $1 \times 1$~km grid cells. Each cell is associated with the static geospatial features described above. We partition time into three-month time intervals, which allows us to capture seasonal trends and corresponds to approximately how often rangers plan new patrol strategies. For each time interval, we aggregate the patrol effort at each cell in terms of distance patrolled in kilometers across each cell. Ideally, we would be able to encode the time spent in each cell to measure patrol effort along with distance covered, as walking slowly through an area likely increases chance of detecting illegal activity. However, rangers typically record waypoints only once every 30 minutes, and activities such as lunch breaks or setting up camp overnight are not identified.

We build the datasets $\mathcal{D} = (\mathbf{X}, \mathbf{y})$ based on these historical patrol observations. The records are discretized into a set of $T$~time steps and $N$~locations to create a matrix $\mathbf{X} \in \mathbb{R}^{T \times N \times k}$, where $k$ is the number of features. Each feature vector~$\mathbf{x}_{t, n}$ contains multiple time-invariant geospatial features associated with each location (described above) and one time-variant covariate: $c_{t-1, n}$, the amount of patrol coverage in cell~$n$ during the previous time step~$t-1$, which models the potential deterrence effect of past patrols. We do not include the current patrol effort $c_{t, n}$ in our feature vector, since we cannot anticipate precisely how much park rangers will patrol in the future when making predictions. 

The labels in the predictive classifier are a binary indicator of whether illegal poaching activity was observed in a cell at a given time step. We assign a positive label $y = 1$ to the cell if rangers observed poaching-related activity during that time period and negative label $y=0$ if they did not. This forms the observation vector $\mathbf{y} \in \{0,1\}^{T \times N}$, where $y_{t, n}$ indicates whether any illegal activity was detected at cell~$n$ during time~$t$.

\begin{figure}
\centering
\input{figures/label_imbalance.tikz}
\caption{Percentage of positive labels at different thresholds of patrol effort. The year in parentheses is used for the test set; the previous three years of data comprise the training set. Note that the $y$-axis varies drastically between datasets.}
\label{fig:label-imbalance-thresholds}
\end{figure}
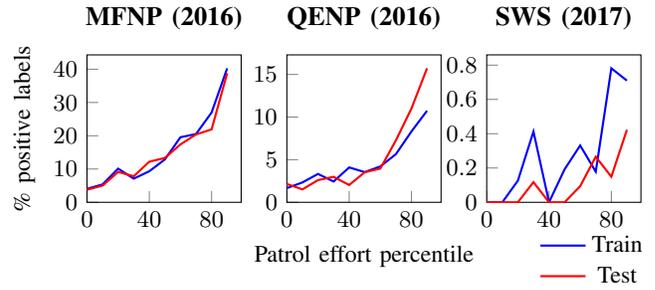

\subsection{Domain Challenges: Uncertainty}

While rangers attempt to record and destroy any signs of illegal activity, they are not always successful---snares may be well-hidden by poachers, and rangers only walk limited paths within each $1 \times 1$~km cell. The success with which they detect poaching activity depends on the amount of effort exerted in patrolling these regions, creating one-sided noise in the observations. Positive records are reliable regardless of the amount of patrol effort (i.e. if rangers find a snare in a cell, poaching occurred with certainty), but negative labels have different levels of uncertainty which depend on the patrol effort~$c_{t,n}$ spent in cell~$n$ during time~$t$. 

As shown in Fig.~\ref{fig:label-imbalance-thresholds}, the percentage of illegal activity detected increases proportionally to patrol effort exerted. Thus, given a threshold~$\theta$ of patrol effort, negative data samples recorded based on a patrol effort of $c_{t,n} \geq \theta$ are relatively more reliable (more likely to be true negatives) compared to the data with less patrol effort $c_{t,n} \leq \theta$. Since negative labels compose a large portion of the wildlife crime datasets, this uncertainty poses a significant challenge to making accurate predictions in this domain.

%% file: figures/label_imbalance.tikz
\begin{tikzpicture}
\pgfplotsset{
  width=0.41\linewidth,
  height=0.4\linewidth,
  xtick pos=left,
  ytick pos=left,
  label style={font=\small},
  tick label style={font=\small},
  xmin=0,
  ymin=0,
}

\begin{axis}[
  at={(0\linewidth,0)},
  title={\textbf{MFNP (2016)}},
  xtick={0,40,80},
  ylabel style={align=center}, 
  ylabel={\% positive labels},
]
\addplot+[thick, mark=none]
  coordinates{
    (0, 4.04312668)
    (10, 5.37153089)
    (20, 10.11638317)
    (30, 7.10382514)
    (40, 9.3250444)
    (50, 12.88014311)
    (60, 19.55156951)
    (70, 20.51971326)
    (80, 26.9955157)
    (90, 40.2690583)
}; 
\addplot+[thick, mark=none]
  coordinates{
    (0, 3.78151261)
    (10, 5.0209205)
    (20, 9.13461538)
    (30, 7.82608696)
    (40, 12.20657277)
    (50, 13.33333333)
    (60, 17.39130435)
    (70, 20.39800995)
    (80, 21.91011236)
    (90, 38.79310345)
}; 
\end{axis}

\begin{axis}[
  at={(0.3\linewidth,0)},
  xlabel={Patrol effort percentile},
  xtick={0,40,80},
  title={\textbf{QENP (2016)}},
]
\addplot+[thick, mark=none]
  coordinates{
    (0, 1.662971175166297)
    (10, 2.3307436182019976)
    (20, 3.3296337402885685)
    (30, 2.436323366555925)
    (40, 4.092920353982301)
    (50, 3.5437430786267994)
    (60, 4.208194905869324)
    (70, 5.647840531561462)
    (80, 8.314855875831485)
    (90, 10.741971207087486)
}; 
\addplot+[thick, mark=none]
  coordinates{
    (0, 2.147239263803681)
    (10, 1.5151515151515151)
    (20, 2.6058631921824107)
    (30, 2.990033222591362)
    (40, 2.027027027027027)
    (50, 3.5294117647058822)
    (60, 3.9568345323741005)
    (70, 7.2607260726072615)
    (80, 11.014492753623188)
    (90, 15.724815724815725)
}; 
\end{axis}

\begin{axis}[
  at={(0.6\linewidth,0)},
  xtick={0,40,80},
  title={\textbf{SWS (2017)}},
  legend style={
    at={(1.1, -.15)},
    legend columns=1, 
    font=\small, 
    draw=none, 
    fill=none
  },
]
\addplot+[thick, mark=none]
  coordinates{
    (0, 0.0)
    (10, 0.0)
    (20, 0.12598425196850394)
    (30, 0.4122766834631242)
    (40, 0.0)
    (50, 0.19241341396371633)
    (60, 0.33140016570008285)
    (70, 0.17857142857142858)
    (80, 0.7815552950371238)
    (90, 0.7095553453169348)
}; \addlegendentry{Train};
\addplot+[thick, mark=none]
  coordinates{
    (0, 0.0)
    (10, 0.0)
    (20, 0.0)
    (30, 0.1176470588235294)
    (40, 0.0)
    (50, 0.0)
    (60, 0.09380863039399624)
    (70, 0.26595744680851063)
    (80, 0.14903129657228018)
    (90, 0.422654268808115)
}; \addlegendentry{Test};
\end{axis}
\end{tikzpicture}

%% file: 4-PredictiveModeling.tex
\section{Predictive Modeling with Uncertain Crime Data}
\label{sec:prediction}

The first stage of PAWS is a predictive model to identify the relative risk of poaching throughout a protected area. We build on the \textbf{i}mperfect observation-a\textbf{ware} \textbf{E}nsemble model (iWare-E) proposed by~\cite{gholami2018adversary}, and enhance this basic iWare-E approach in three ways. 

\begin{figure}
\centering
\includegraphics[width=0.95\columnwidth]{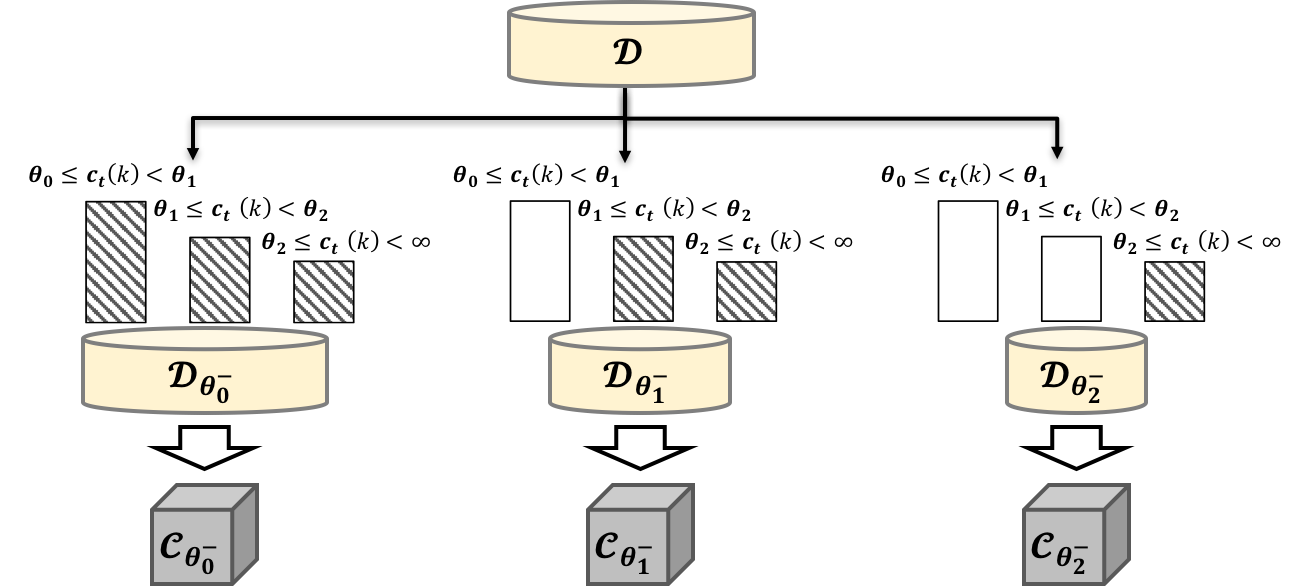}
\caption{The iWare-E model proposed in~\cite{gholami2018adversary}. $\mathcal{D}$~is the entire training dataset, which is modified into datasets $\mathcal{D}_{\theta_i^-}$, represents a subset of the data where negative samples recorded by a patrol effort of $c_{t,n} \leq \theta_i$ (shown by white bars) are removed. This figure shows $I = 3$ weak learners. Each classifier~$\mathcal{C}_{\theta_i^-}$ is trained on a subset~$\mathcal{D}_{\theta_i^-}$. The remaining negative labels are shown by shaded bars.} 
\label{fig:iWare-E}
\end{figure}

The iWare-E model, diagrammed in Fig.~\ref{fig:iWare-E}, generates subsets of the data by filtering patrol effort across different thresholds to produce an ensemble of learners. Each weak learner~$\mathcal{C}_{\theta_i^-}$ is trained on a subset of the dataset~$\mathcal{D}_{\theta_i^-}$, filtered by a threshold~$\theta_i$ for $i \in \lbrace 1, \ldots, I \rbrace$ such that $\theta_i \leq \theta_{i+1}$. For any choice of $[\theta_{\text{min}}$, $\theta_{\text{max}}]$, $I$ intermediate thresholds $\theta_i$ can be obtained such that ${\theta_{\text{min}} \leq \theta_i \leq \theta_{\text{max}}}$. Due to the label imbalance, we discard only negative samples and keep all positive samples, even those below the specified threshold; this is one of the key insights of the iWare-E approach. Thus, we train $I$ weak learners based on those subsets of the data. \cite{gholami2018adversary}~used a bagging ensemble of either decision trees or SVM models as the weak learners.


First, to improve on the iWare-E ensemble method, we introduce a strategic approach to computing classifier weights. In iWare-E, we build $I$ classifiers, each trained on data with patrol effort of threshold~$\theta_i$. \cite{gholami2018adversary} proposed that each classifier~$C_{\theta_i^-}$ is qualified to make predictions on test data with threshold $\theta_j \leq \theta_i$ and that the qualified predictions should be weighed equally. With that approach, classifiers trained on data with high patrol effort would only predict on data with equal or higher patrol effort. We instead propose a systematic way to compute optimal classifier weights. When training, we hold out a testing set and perform 5-fold cross validation to minimize the log loss of the predictions when varying the classifier weights. We then retrain the classifiers across the entire training data, with these optimal weights computed. 





Second, we streamline the process of selecting patrol thresholds to build the various classifiers in iWare-E. In picking thresholds~$\theta_i$ for patrol effort, \cite{gholami2018adversary}~used 16 equally-distanced values from $\theta_1 = 0$ km to $\theta_I = 7.5$ km. However, we found that the best approach was to select these thresholds based on patrol effort percentiles, to produce a consistent amount of training data for each classifier. This approach then simplifies the model such that the number of classifiers becomes a single hyperparameter, rather than having to pick $\theta_{\text{min}}$, $\theta_{\text{max}}$, and $\Delta \theta$ separately. The number of classifiers should be selected based on the characteristics of the data; we used more classifiers (20) for QENP and MFNP, which have well-behaved label imbalance (see Fig.~\ref{fig:label-imbalance-thresholds}), than for SWS (10). In addition, selecting thresholds based on percentile better accounts for sparsity in the data. For example, there may be very few cells patrolled with effort between 5 and 6~km, and those points may all be negative labels.


\label{sec:GPuncertainty}
Finally, a major enhancement we introduce is enhancing the iWare-E model to explicitly reason about uncertainty of the predictions. We augment the iWare-E method with Gaussian process~(GP) classifiers~\cite{rasmussen2004gaussian} as the weak learners. GPs are given by the function:
\begin{equation}
f(\mathbf{x}_i)\sim \mathcal{GP} \left( \mu(\mathbf{X}), \Sigma(\mathbf{X}) \right) \ ,
\end{equation}
with mean~$\mu(\mathbf{X})$ and covariance matrix~$\Sigma(\mathbf{X})$. Due to the formal definition of the covariance functions, GPs compute a variance value associated with each prediction based on confidence from the training data. We use an implementation of Gaussian process classifiers from~\cite{rasmussen2004gaussian}, which computes variance values to quantify the uncertainty of each prediction. Later on, in the patrol planning stage, we make use of these variance values as the metric for uncertainty in order to plan more informed patrol routes.

%% file: 5-EvaluatePredictiveModeling.tex
\section{Evaluation of the Predictive Model}
We study the predictive performance across different datasets by varying the weak learner used in the iWare-E model, showing that our enhanced iWare-E approach raises AUC by 0.100 on average compared to a bagging weak learner, and that iWare-E with GPs shows the most consistently strong performance compared to SVMs or decision trees. We perform in-depth analysis of the uncertainty values from Gaussian processes and compare those values with a heuristic metric for uncertainty from bagging decision trees. 

\subsection{Evaluation on Historical Data}

We generate predictive poaching models with four years of data for each park, training on the first three years and testing on the fourth. This setup simulates the ability of each model to predict future incidences of poaching. Although we have up to 18~years of data for each park, earlier years are increasingly less predictive of future years: a park may hire more rangers; the sale price of illegal wildlife goods may increase, making poaching more attractive; or logging may increase in a region, thus changing the landscape. 

\begin{table}
\centering
\caption{About the datasets}
\label{table:datasets}
\begin{tabular}{l|cccc}
\Xhline{2\arrayrulewidth}
 & \textbf{MFNP} & \textbf{QENP} & \textbf{SWS} & \textbf{SWS dry} \\
\hline
Number of features & 22 & 19 & 21 & 21 \\
Number of $1 \times 1$~km cells & 4,613 & 2,522 & 3,750 & 3,750 \\
Number of points (6 years) & 18,254 & 19,864 & 43,269 & 30,569 \\
Number of positive labels & 2,602 & 937 & 155 & 76 \\
Percent positive labels & 14.3\% & 4.7\% & 0.36\% & 0.25\% \\
Avg. patrol effort (km/cell) & 1.75 & 2.08 & 3.96 & 3.03 \\
\Xhline{2\arrayrulewidth}
\end{tabular}
\end{table}

The dataset from SWS introduced new challenges of class imbalance. Observe in Table~\ref{table:datasets} the extreme label imbalance in SWS, with only 0.36\% positive labels. Given these new challenges of significantly imbalanced data, we used a balanced bagging classifier to undersample negative labels~\cite{he2008learning}, implemented by \cite{guillaume2017imblearn}. This undersampling approach improved our AUC by 15\% on average on the SWS dataset. With wildlife crime data, undersampling is preferred to oversampling the minority class because the positive labels are inherently noisy due to random factors that influence whether rangers detect poaching activity. 

We implement the iWare-E ensemble method with our enhancements described above, using three base classifiers: bagging ensembles of SVMs~(SVB-iW), bagging ensembles of decision trees~(DTB-iW), and bagging ensembles of Gaussian process classifiers~(GPB-iW). We compare these models to baseline models, using those same bagging weak learners but without iWare-E (referred to as SVB, DTB, and GPB).

\newcolumntype{g}{>{\columncolor{tablegray}}c}
\begin{table}
\centering
\caption{Comparing performance (AUC) of each model across all datasets}
\label{table:metrics}
\begin{tabular}{cg|ggg|ggg}
\Xhline{2\arrayrulewidth}
\rowcolor{white} & & \multicolumn{3}{c|}{\small{\textbf{without iWare-E}}} & \multicolumn{3}{c}{\small{\textbf{with iWare-E}}} \\
\rowcolor{white} & & \textbf{SVB} & \textbf{DTB} & \textbf{GPB} & \textbf{SVB} & \textbf{DTB} & \textbf{GPB} \\
\hline
\rowcolor{white} & \textbf{2014} & 0.518 & 0.587 & 0.626 & 0.695 & \textbf{0.711} & 0.685 \\
\rowcolor{white} & \textbf{2015} & 0.504 & 0.620 & 0.670 & 0.683 & 0.706 & \textbf{0.726} \\
\rowcolor{white} & \textbf{2016} & 0.513 & 0.589 & 0.603 & 0.672 & 0.680 & \textbf{0.681} \\ 
\multirow{-4}{*}{\rotatebox[origin=c]{90}{\textbf{MFNP}}} 
& \textbf{Avg} & 0.512 & 0.599 & 0.633 & 0.683 & \textbf{0.699} & 0.697 \\ \hline
\rowcolor{white} & \textbf{2014} & 0.505 & 0.654 & 0.693 & 0.619 & \textbf{0.735} & 0.717 \\
\rowcolor{white} & \textbf{2015} & 0.501 & 0.589 & 0.600 & 0.632 & 0.696 & \textbf{0.713} \\
\rowcolor{white} & \textbf{2016} & 0.502 & 0.635 & 0.611 & 0.644 & 0.728 & \textbf{0.733} \\ 
\multirow{-4}{*}{\rotatebox[origin=c]{90}{\textbf{QENP}}} 
& \textbf{Avg} & 0.503 & 0.626 & 0.635 & 0.632 & 0.720 & \textbf{0.721} \\ \hline
\rowcolor{white} & \textbf{2016} & 0.815 & 0.774 & 0.656 & \textbf{0.870} & 0.865 & 0.825 \\
\rowcolor{white} & \textbf{2017} & 0.672 & 0.670 & 0.673 & 0.744 & 0.742 & \textbf{0.847} \\
\rowcolor{white} & \textbf{2018} & 0.518 & 0.499 & 0.527 & 0.510 & 0.541 & \textbf{0.680} \\
\multirow{-4}{*}{\rotatebox[origin=c]{90}{\textbf{SWS}}}
& \textbf{Avg} & 0.668 & 0.648 & 0.619 & 0.708 & 0.716 & \textbf{0.784} \\ \hline
\rowcolor{white} & \textbf{2016} & 0.500 & 0.490 & 0.648 & 0.501 & 0.610 & \textbf{0.771} \\
\rowcolor{white} & \textbf{2017} & 0.500 & 0.486 & 0.615 & 0.581 & 0.681 & \textbf{0.827} \\
\rowcolor{white} & \textbf{2018} & 0.500 & 0.526 & 0.615 & 0.500 & 0.576 & \textbf{0.674} \\
\multirow{-4}{*}{\rotatebox[origin=c]{90}{\textbf{SWS dry}}}
& \textbf{Avg} & 0.500 & 0.501 & 0.626 & 0.527 & 0.622 & \textbf{0.757} \\
\Xhline{2\arrayrulewidth}
\end{tabular}
\end{table}

Table~\ref{table:metrics} presents the performance of the different predictive models evaluated on MFNP, QENP, and SWS for various choices of weak learners. The year listed is the test set; for example, MFNP~(2016) indicates that 2013--2015 were used for training and 2016 for testing. We conducted three experiments on each of the parks: with test sets 2014, 2015, and 2016 for MFNP and QENP, and 2016, 2017, and 2018 for SWS. In addition, since SWS experiences significant seasonality, we also process SWS using data from just the dry season (November--April), as used for the field tests in Section~\ref{sec:sws_field_test}. Note that to process dry season, we discretize time into two-month periods (rather than three) to obtain three points per year. 

The iWare-E approach consistently improves AUC across all models, raising the AUC by 0.100 on average. Specifically, the GPB-iW model yields the best performance in over half the cases. Note that GPs (both with and without iWare-E) achieve significant performance gains in SWS dry season, which has the strongest class imbalance with only 0.25\% positive labels, as listed in Table~\ref{table:datasets}. In general, SVMs are suboptimal weak learners in this domain, and decision trees and GPs have comparable performance. Thus, introducing GPs to iWare-E does not inhibit performance in any cases, and improves performance in cases of strong class imbalance. As we will show, the ability of GPs to quantify uncertainty provides advantages in patrol planning that make it the obvious preferred weak learner.

\subsection{Analysis of the Uncertainty in Predictions}

\label{sec:predictionUncertainty}
\begin{figure*}
\centering
\begin{subfigure}{.2\textwidth}
\centering
\includegraphics[width=\textwidth]{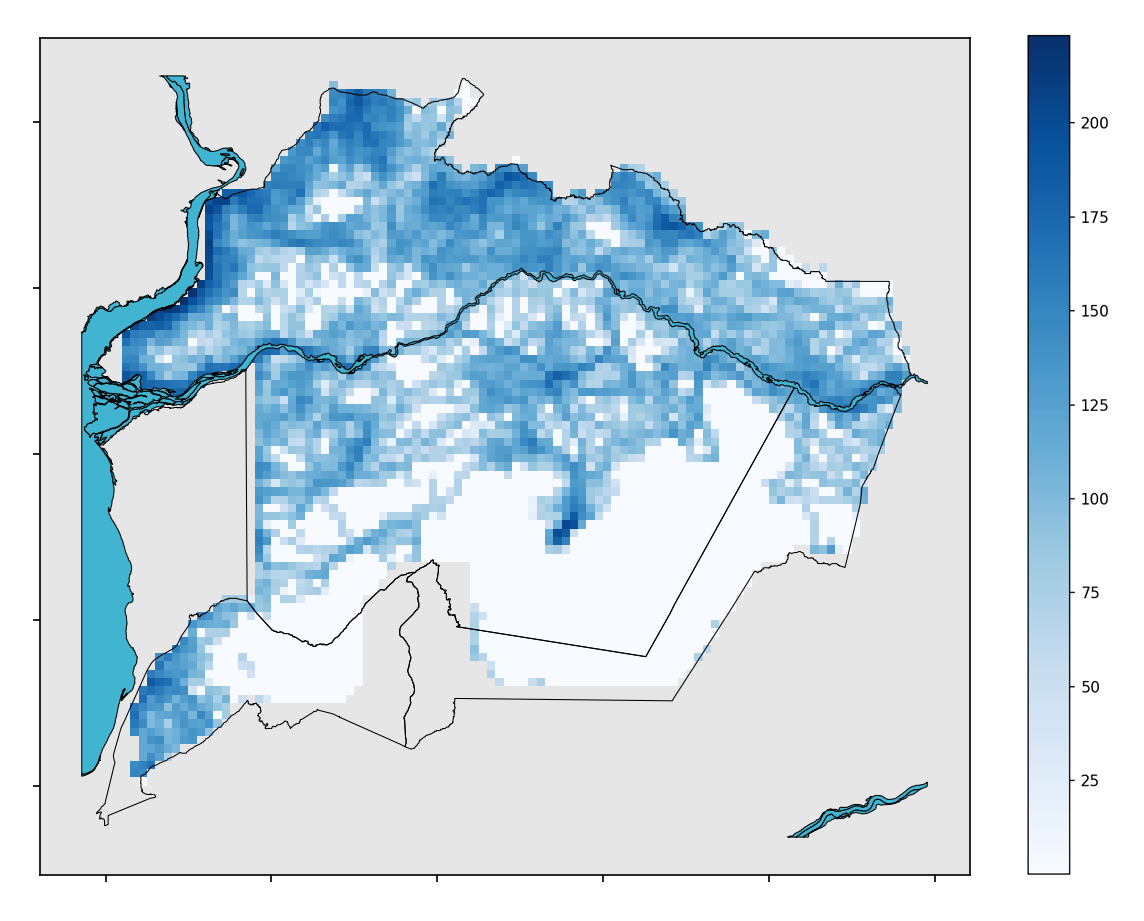}
\caption{Historical patrol effort from 2014--2016.}
\label{fig:MFNPeffort}
\end{subfigure}
\begin{subfigure}{.2\textwidth}
\centering
\includegraphics[width=\textwidth]{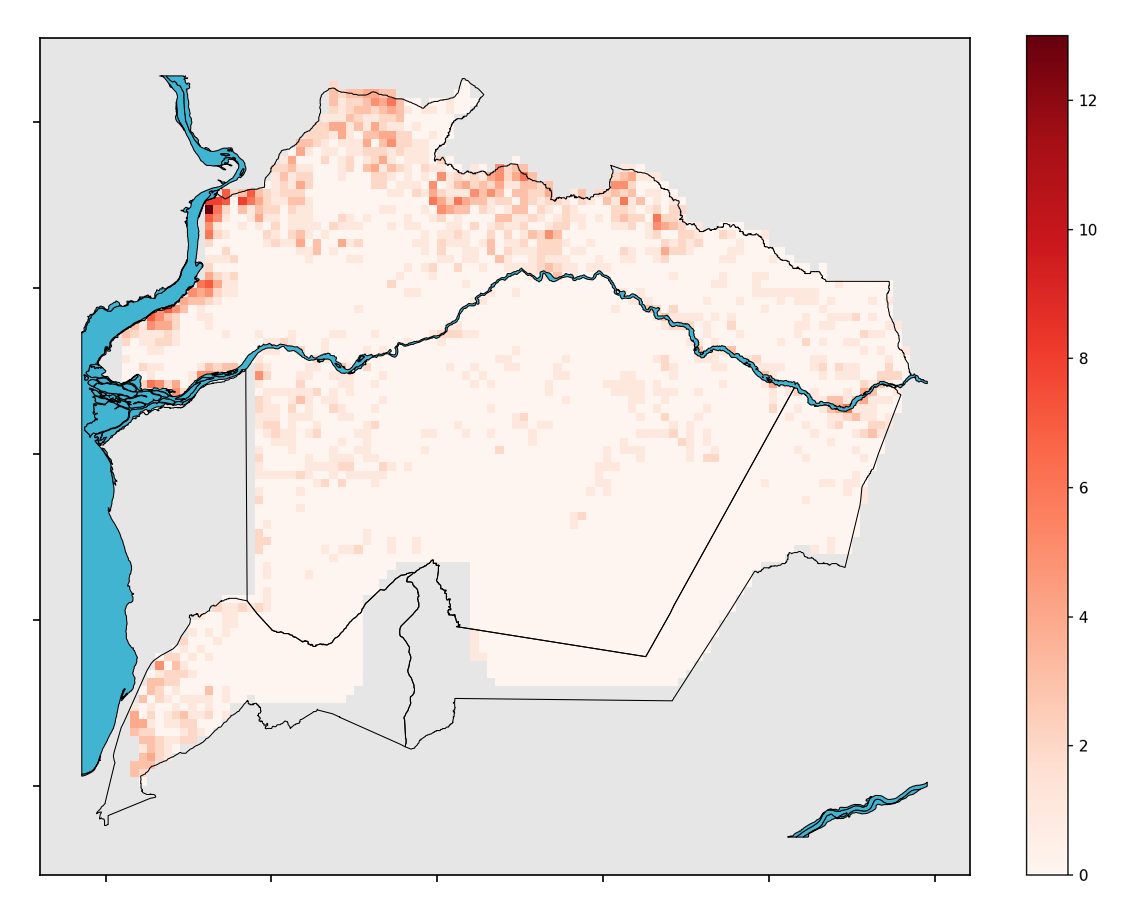}
\caption{Historical illegal activity detected over 2014--2016.}
\label{fig:MFNPactivity}
\end{subfigure}
\qquad
\begin{subfigure}{.52\textwidth}
\centering
\includegraphics[width=\textwidth]{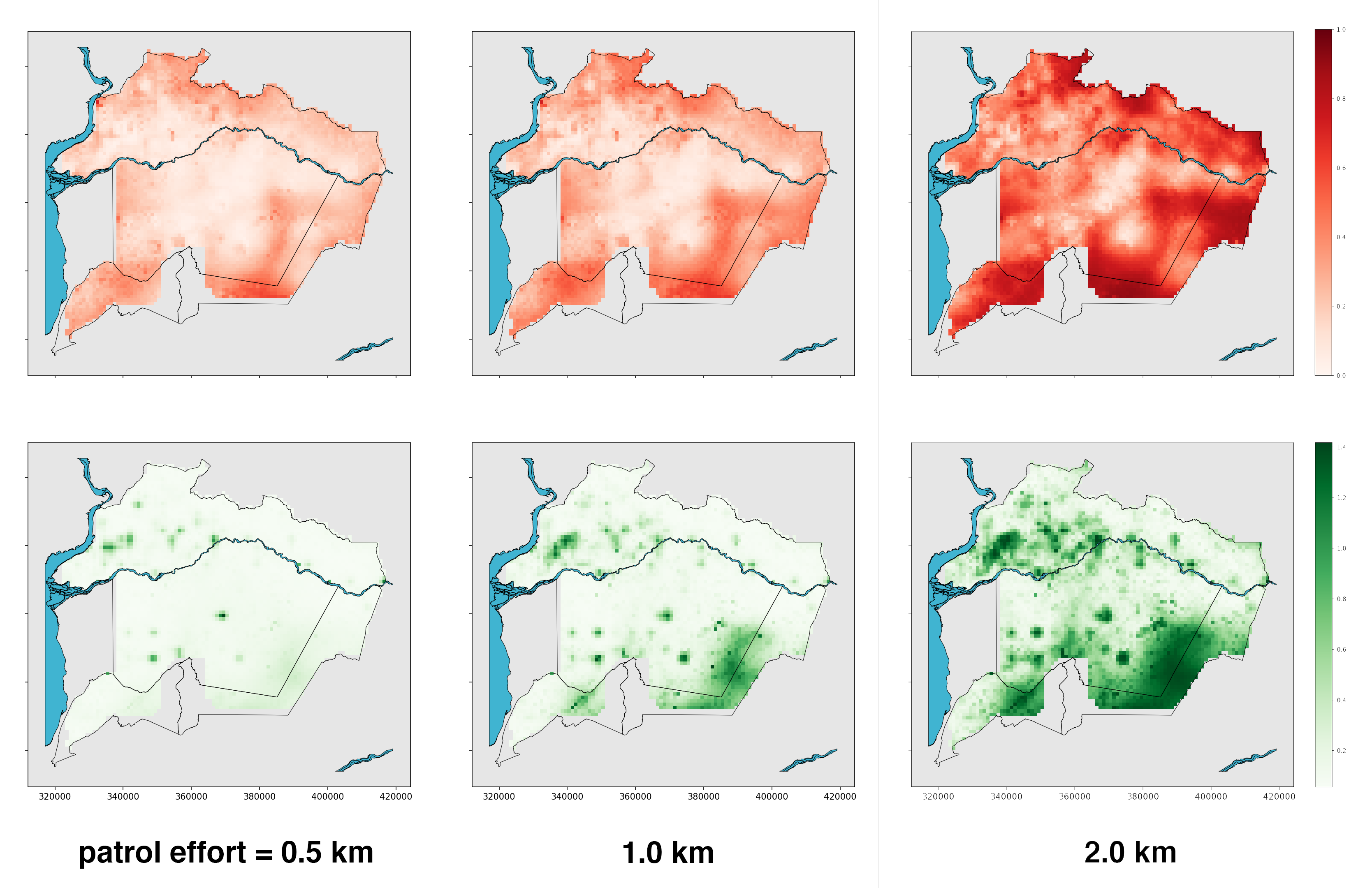}
\caption{Predicted probability of detecting of poaching activity in 2017 (above, red) and corresponding uncertainty of the predictions (below, green).}
\label{fig:pred}
\end{subfigure}
\caption{Prediction values and uncertainties for different levels of patrol effort in MFNP in Uganda in the first quarter of 2017.}
\label{fig:prediction_uncertainty}
\end{figure*}

We analyze the predicted probabilities and the corresponding uncertainties in Murchison Falls National Park as computed by the GPB-iW model, shown in Fig.~\ref{fig:prediction_uncertainty}. The riskmaps and uncertainty maps are displayed along with the historical patrol data over three years: total distance (km) that rangers patrolled in each cell (Fig~\ref{fig:MFNPeffort}), and the number of incidences of illegal activity detected during those patrols (Fig~\ref{fig:MFNPactivity}). 

Fig.~\ref{fig:pred} (red) depicts the predicted risk of detecting illegal activity in a cell, which is the joint probability that poachers attack a cell and that rangers also observe the attack $\Pr[a=1, o=1]$. These predictions are generated by GPB-iW across different levels of patrol effort in MFNP. For instance, the first plot shows the predicted probability of detecting an attack if each 1 sq. km area in the park is patrolled with 0.5~km of patrol effort by rangers during three months of patrolling. The predicted probability generally increases as patrol effort increases, although the increase is not uniform. Some of these cells have near-zero probability ($\Pr[a=1, o=1] \simeq 0$) despite increased patrol effort, indicating almost no likelihood of attack ($\Pr[a=1] \simeq 0$) in those cells. Regions where the predicted probability increases slowly signify that likelihood of detection plateaus, and it would be wasteful to allocate excessive patrolling resources in those areas. Observe that several areas with historically high levels of patrol effort, such as the northwest section, are predicted to have low risk of detecting poaching activity. This finding that rangers should instead focus on patrolling elsewhere. 

These uncertainty heatmaps in Fig.~\ref{fig:pred} (green) show the model's confidence of each prediction. For example, the southeast region of the park suffers from the highest amount of uncertainty in the predictions. As expected, the historical patrol effort (Fig.~\ref{fig:MFNPeffort}) reveals that patrolling has been minimal in this region, due in part to few large mammals and lack of patrol posts. Since this region has the least amount of historical data, we have the greatest uncertainty in the predictions due to lack of prior observations exposed to the predictive model. This information is valuable for robust defender patrol planning which strategizes against the worst-case attack scenarios, and could also be used to plan patrol routes that explicitly target areas with high model uncertainty in order to reduce the existing data bias. 
Observe that uncertainty increases as we make predictions with high levels of patrol effort, which is explained by the fact that historical data with higher levels of patrol effort is more rare.

\subsection{Comparison of Uncertainty in Gaussian Processes and Bagging Decision Trees}

The bagging ensemble of decision trees we use is equivalent to a random forest, as the bagged learners vary in both data and features sampled. The confidence interval of a random forest is defined as the variance between predictions made by the bagged learners, which can be used as a heuristic for uncertainty\cite{mentch2016quantifying}. We compute confidence intervals using the infinite jackknife method proposed by~\cite{wager2014confidence} to compare against the variance computed by Gaussian processes. 

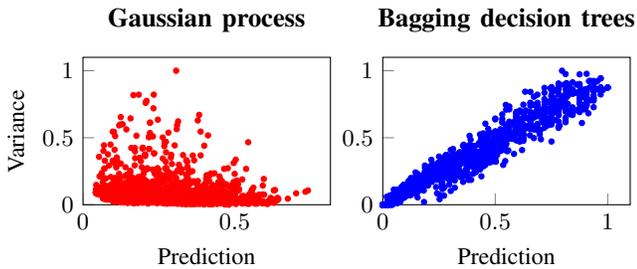
\begin{figure}
\input{figures/correlation_plots.tikz}
\caption{Correlation plots for prediction vs. uncertainty (variance) in Gaussian processes and bagging decision trees. Data shown is one classifier $\mathcal{C}_{\theta_i^-}$ run on MFNP~2016.}
\label{fig:correlation_plots}
\end{figure}

Fig.~\ref{fig:correlation_plots} visualizes the correlation plots for one classifier run on the MFNP dataset, with test year 2016. The Pearson correlation coefficient is $-0.198$ for GPs, but $0.979$ for bagging decision trees---a near-perfect correlation. Thus, the variance values for bagging decision trees provides little additional insight. In addition, the variance value in Gaussian processes is an actual metric intrinsic to the model, whereas in random forests the confidence interval is just a surrogate metric. Thus, we cannot substitute the variance metric from bagging decision trees as uncertainty; GPs are necessary for this insight. 

%% file: figures/correlation_plots.tikz
\begin{tikzpicture}
\pgfplotsset{
  width=0.55\linewidth,
  height=0.4\linewidth,
  xtick pos=left,
  ytick pos=left,
  label style={font=\small},
  tick label style={font=\small},
  xmin=0,
  ymin=0,
}

\begin{axis}[
  at={(0\linewidth,0)},
  title={\textbf{Gaussian process}},
  xlabel={Prediction},
  ylabel={Variance},
  xtick={0,.5,1},
]
\addplot[
    only marks,
    color=red,
    mark size=1pt,
    mark=*,]
  table{figures/correlation_plot_gp.dat};
\end{axis}

\begin{axis}[
  at={(0.45\linewidth,0)},
  title={\textbf{Bagging decision trees}},
  xlabel={Prediction},
  xtick={0,.5,1},
]
\addplot[
    color=blue,
    only marks,
    mark size=1pt]
  table{figures/correlation_plot_rf.dat};
\end{axis}
\end{tikzpicture}

%% file: 7-PatrolPlanning.tex
\section{Patrol Planning}
Having assessed the relative risk of poaching activity throughout a protected area with our predictive model, we then employ a game-theoretic model to plan optimal patrol routes. In this section, we enhance the patrol planning model proposed in \cite{gholami2018adversary} by incorporating uncertainty in the predictions of the poacher's behavior, a metric generated by GPs. Conducting risk-adverse patrols enables us to increase detection of snares by an average of 30\%. 


\subsection{Game Model}
We use the game model from \cite{gholami2018adversary}, which is summarized here for completeness. We discretize a protected area into a set of $N$~grid cells (corresponding to $1 \times 1$ km regions) to form a graph~$G=(V, E)$ of nodes and edges. We model the problem of allocating park patrols as a game on this graph, played between a defender (rangers) and a set of $N$ adversaries (poachers), located at each cell. Each adversary may choose to attack their cell by placing snares to catch animals. The rangers attempt to thwart these attacks by conducting patrols to detect snares, and receive a payoff of $1$ for every attack successfully thwarted. Rangers conduct their patrols in the conservation area over a period of $T$ time steps, where a single time step corresponds to the minimum amount of time required to cross one cell; that is, rangers must spend at least one time step in each cell they choose to visit. A ranger's patrol is then a path taken on a time-unrolled graph $G'=(V', E')$, with a set of nodes indicating location and time. The set of possible paths on the time-unrolled graph forms the pure strategy space of the defender.

Each patrol must begin and end at nodes in the graph designated as patrol posts, which we refer to as the source $s \in V$ for graph~$G$. In the time-unrolled graph we designate the source of any patrol as $s_1 = (s, 1) \in V'$ and the target $s_T = (s,T) \in V'$ corresponding to the patrol post visited in the first and last time steps. The ability of rangers to detect poaching in any cell $v \in V$ depends on the level of patrol effort~$c_{t,v}$ in that cell at time~$t$. Rangers may increase the level of patrol effort at any cell by either spending multiple time steps at the cell in a single patrol or visiting that cell during multiple distinct patrols. We specify a set of constraints, so that a single feasible patrol corresponds to one unit of flow on the time-unrolled graph; the sum of the total flow over all edges in $G'$ is then equal to~$T$. The pure strategy space is the set of such flows~$\mathcal F$, given by
\begin{small}
\begin{align}
\mathcal{F} :=
\left\{ 
f_{u',v'} :
\begin{array}{ll}
\sum\limits_{u' : (u',v')\in E'} f_{u',v'} = \sum\limits_{u' : (v',u')\in E'} f_{v',u'} \;\;\forall v' \in N'\\
\sum\limits_{u': (s_1,u')\in E'} f_{s_1,u'} = \sum\limits_{u': (u',s_T)\in E'} f_{u',s_T} = 1\\
\sum\limits_{(v',u')\in E'} f_{u',v'} = T \\
\end{array}
\right\}
\end{align}
\end{small}

Let $x_v$ be the defender coverage at cell~$v$, which also represents the defender mixed strategy. Let $a_v$ be the action of the adversary located at cell~$v$. Each adversary responds to the mixed strategy coverage~$c_v$ of the defender at that cell and chooses at the beginning of the game either to attack $a_v(x_v)=A$ or not attack $a_v(x_v)=\neg A$ with some probability $\Pr[a_v(x_v)]$ such that $ \Pr[a_v(x_v) = A] + \Pr[a_v(x_v) = \neg A] = 1$. The defender utility is conditioned on a successful detection of attack. Let $o_v(x_v) = O$ denote a successful detection of snares at cell~$v$. Note that the detection success probability is also a function of the defender's mixed strategy coverage. The defender expected utility~$U^d$ is then the probability of detecting snares at a cell, given that there is an attack at the cell, summed over all cells
\begin{align}
    U^d(a,x) = \sum_{v \in N} \Pr[o_v=O \mid a_v = A] \Pr[a_v=A] \ .
\end{align}
Note that we have omitted the dependence on $x_v$ for ease of notation. We compute an equilibrium solution to this game where the defender attempts to maximize her utility function $U^d(a,x)$ and each of the adversaries maximizes their own utility function $U^a_v(a_v, x_v)$. If the adversaries were perfectly rational, this would correspond to a strong Stackelberg equilibrium with each of the adversaries as in standard Stackelberg security games. However, in Green Security Games, the adversaries are boundedly rational and we learn adversary behavior models from data, as described in Sec.~\ref{sec:prediction}. Thus, we achieve equilibrium with non-rational attackers as in~\cite{nguyen2016capture,yang2014adaptive}.

\subsection{Prescriptive Modeling with Certain Crime Predictions}

The adversary utility and the probability of detecting snares given that the adversary chooses to attack are both unknown functions. We use past data to learn a predictive model of the adversary's response to the defender mixed strategy~$x$ as well as the defender detection probability. To compute solutions to the patrol planning game, we then need to optimize this predictive model, which we treat as a black-box function. We use a similar framework as in \cite{gholami2018adversary} to perform this optimization, which allows us to not only plan with a black-box objective function, but also to reason about continuous decision variables such as patrol effort. The model uses a piecewise linear objective function to approximate predictions of the model. This formulation was shown to have significant improvements in runtime compared to previous methods for patrol planning with a black-box function, which could only reason about discrete levels of patrol effort.

The predictive model produces for each cell a function $g_v : c_v \rightarrow P_v$ which maps the total defender patrol effort at a cell $v \in N$ to a corresponding likelihood that there will be a detected attack~$P_v$ at that cell. The defender patrol effort is a function of the defender mixed strategy \begin{small}$c_v = x_v K$\end{small} where $K$ is the number of patrols that the defender conducts. As in~\cite{gholami2017taking}, piecewise linear (PWL) approximations to these functions~$g_v$ are constructed using $m \times N$ sampled points from the $N$ functions~$g_v$. 

These define the optimization problem~(\ref{planning_prob}) which can be expressed as a mixed integer linear program~(MILP):
\begin{align*}
\max_{c, f} & \sum_{v\in N}g^{\text{PWL}}_v(c_{v}) \\
& f_{u',v'} \in \mathcal F & \forall (u',v') \in E' \\
& K \sum_{u': (v', u') \in E'} f_{u', v'} = c_v & \forall v \in N, v' = (v,t) \\
& \sum_{v \in N} c_{v} = T \times K
\label{planning_prob}
\tag{$\mathcal{P}$}
\end{align*}
Our objective function is given by the PWL approximation to the machine learning model predictions, which we refer to as~$g_v^{\text{PWL}}$. The first constraints are the flow constraints on the time unrolled graph $G=(N',E')$. The second constraint enforces that the patrol effort is equal to the amount of flow into a node~$v$ across all time (which gives the defender mixed strategy coverage~$x_v$) times the number of patrols~$K$. The last constraint enforces that the total patrol effort expended is equal to the length of each patrol~$T$ times the number of patrols~$K$.

\subsection{Prescriptive Modeling with Uncertain Crime Predictions}
We incorporate predictions from the GPB-iW model to plan patrol routes for rangers. We only have black-box access to the predictions as a function of the rangers patrol effort, which we need to be able to optimize in order to compute patrol paths. The planning model in~\cite{gholami2018adversary} allows us to create paths, but does not account for uncertainty in the predictions or have a game theoretic component. To account for this, we augment the model by using the variance associated with each of the predictions from the GPB-iW model to plan patrol routes.

The new GPB-iW machine learning model now gives for each cell~$v$ a variance function $\nu_v : c_v \rightarrow \mathcal{V}_v$, where $\mathcal{V}_v$ is the variance at cell~$v$ that becomes the uncertainty score for each prediction~$g_v(c_v)$, with likelihood~$P_v$ that there will be a detected attack at that cell. We want to compute a series of patrols which not only maximize the probability of detecting attacks over the entire area, but also takes into account the uncertainty of each prediction. To do this, we take a robust approach by penalizing the expected probability of detection~$g_v$ by a scaled function of the uncertainty score~$\nu_v$. The utility of patrolling any cell~$v$ as a function of the patrol effort~$c_v$ is:
\begin{equation}
U_v(c_v) = g_v(c_v) - \beta g_v(c_v) \nu_v(c_v) \ .
\end{equation}

The uncertainty scores that we get from the GPB-iW model are scaled to the range $[0,1]$ through a logistic squashing function. We then choose $\beta \in [0,1]$ to rescale the uncertainty score and ensure that the objective function is always positive. Larger values of~$\beta$ allow us to optimize for plans which are more risk averse; $\beta$ thus becomes a parameter that enables us to tune the robustness of our approach. $\beta > 0$ corresponds to a pessimistic, risk-averse approach where we will patrol less in cells where there is greater uncertainty. We can compute the optimal patrol by substituting 
\begin{align}
    \sum_{v\in N}U^{\text{PWL}}_v( c_{v})= \sum_{v\in N} \left( g^{\text{PWL}}_v(c_v) - \beta g^{\text{PWL}}_v(c_v) \nu_v^{\text{PWL}}(c_v) \right)
\end{align} 
as our objective function in $\mathcal{P}$, where, as in~\cite{gholami2018adversary} we construct PWL approximations to $g_v$ and~$\nu_v$ so that the optimization problem is expressible as a MILP.

\subsection{Evaluation of the Prescriptive Model}

\begin{figure}
\input{figures/improvement.tikz}
\caption{Improvement in detection of snares when accounting for uncertainty in patrol planning. Fig. (a)--(c) show the improvement in solution quality measured by the ratio $U_{\beta}(C_{\beta}) / U_{\beta}(C_{\beta=0})$ as a function of the tuning parameter~$\beta$ which determines the robustness of the solutions through the weight on the uncertainty score. Fig. (d)--(f) show improvement in solution quality with increasing segments in the PWL approximation to the GPB-iW predictions.}
\label{robust}
\end{figure}
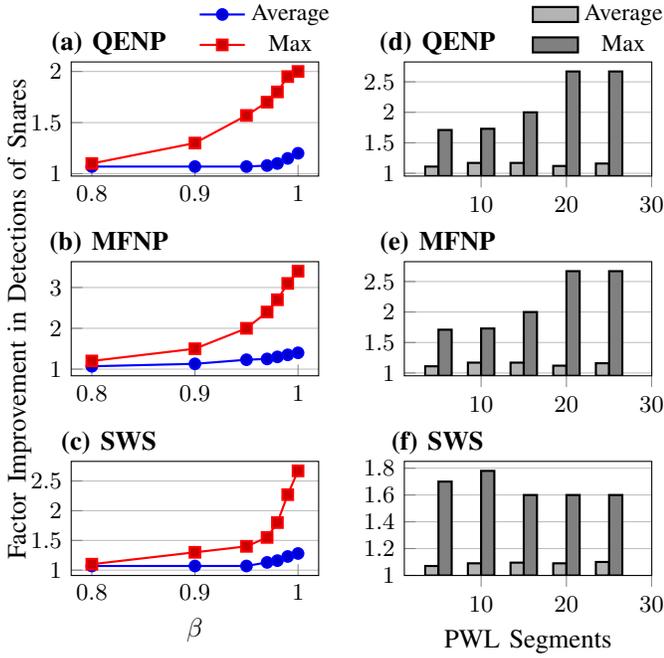

\begin{figure}[htb]
\input{figures/pwl_functions.tikz}
\caption{Prescriptive model runtime (a) and patrol plan utility (b) as a function of the number of segments in the PWL function approximation to the GPG-iW model.}
\label{PWL_util}
\end{figure}
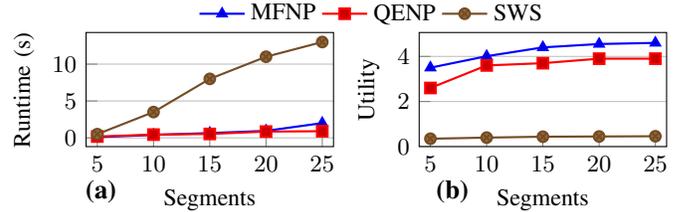

To demonstrate the benefit of accounting for uncertainty when planning patrols, we compare the patrols computed with and without uncertainty scores by evaluating them on the ground truth given by the objective with uncertainty. We refer to the plan computed using uncertainty weighted with value $\beta$ as $C_{\beta} = \text{argmax}_c \sum_v g_v(c_v)-\beta g_v(c_v)\nu_v(c)$, such that $C_{\beta=0}$ is a plan which does not account for uncertainty and $C_{\beta=1}$ is a fully robust plan. We then evaluate each of the plans using a utility function $U_{\beta}(C)$ and compute the ratio of the solution quality of the plan at a given $\beta$ to the baseline of $\beta=0$, $U_{\beta}(C_{\beta}) / U_{\beta}(C_{\beta=0})$. For each dataset, we looked at the gain in solution quality for plans generated for each patrol post in the park, evaluated by varying the $\beta$~parameter as well as the number of segments in the piecewise linear functional approximations to~$U_v$. The results are shown in Fig.~\ref{robust}, where we consider both the average gain over all patrol posts as well as the maximum gain achieved. Not only does this demonstrate the benefit of robust planning, but rangers may also use this knowledge to help decide where to collect more data. By looking at the ratio $U_{\beta}(C_{\beta}) / U_{\beta}(C_{\beta=0})$ we can determine whether reducing the uncertainty in the predictive model will affect the computed plans. When this ratio is close to $1$ there is little benefit in collecting data in order to reduce uncertainty in predictions as it does not change the utility of the computed patrols. 
However, when the ratio is large, the computed patrols are highly dependent on the areas where the model is uncertain; therefore there can be significant gains in utility through the reduction of uncertainty through collecting more data. 
We also demonstrate the scalability of this approach in Fig.~\ref{PWL_util}a, which shows the runtime with increasing number of breakpoints in the PWL function approximation. Fig.~\ref{PWL_util}b plots the convergence of the utility of the computed solutions. We looked at the utility of the robust solutions $U_{\beta=1}(C_{\beta=1})$ with increasing number of segments in the PWL function and see that utility of the optimal solution converges around 20--25 segments. 

%% file: figures/improvement.tikz
\begin{tikzpicture}
\pgfplotsset{
  width=0.55\linewidth,
  height=0.35\linewidth,
  xtick pos=left,
  ytick pos=left,
  tick label style={font=\small},
  ymajorgrids=true,
}
\begin{axis}[ 
  at={(0, 0.6\linewidth)},
  bar width=5,
  legend style={
    at={(1.1,1.6)},
    legend columns=1,
    font=\small, 
    draw=none,
    fill=none},
]
\addplot+[thick]
  coordinates{
    (0.8,1.07)
    (0.9,1.07)
    (0.95,1.07)
    (0.97,1.08)
    (0.98,1.1)
    (0.99,1.15)
    (1,1.2)
}; \addlegendentry{Average};
\addplot+[ thick]
  coordinates{
    (0.8,1.1)
    (0.9,1.3)
    (0.95,1.57)
    (0.97,1.7)
    (0.98,1.8)
    (0.99,1.95)
    (1,2.0)
}; \addlegendentry{Max};
\end{axis}
\begin{axis}[
  at={(0, 0.3\linewidth)},
  bar width=5,
  ylabel style={align=center}, 
  ylabel={Factor Improvement in Detections of Snares},
]
\addplot+[thick,area legend]
  coordinates{
   (0.8,1.07)
    (0.9,1.13)
    (0.95,1.23)
    (0.97,1.25)
    (0.98,1.3)
    (0.99,1.35)
    (1,1.4)

}; 
\addplot+[ thick,area legend]
  coordinates{
    (0.8,1.2)
    (0.9,1.5)
    (0.95,2)
     (0.97,2.4)
    (0.98,2.7)
    (0.99,3.1)
    (1,3.4)
}; 
\end{axis}
\begin{axis}[
  at={(0, 0\linewidth)},
  xlabel=$\beta$, 
  bar width=5,
  legend style={at={(1.1,1.35)},legend columns=2,font=\small, draw=none,fill=none},
]
\addplot+[thick]
  coordinates{
    (0.8,1.07)
    (0.9,1.07)
    (0.95,1.07)
    (0.97,1.13)
    (0.98,1.16)
    (0.99,1.23)
    (1,1.28)
}; 
\addplot+[ thick]
  coordinates{
    (0.8,1.1)
    (0.9,1.3)
    (0.95,1.4)
    (0.97,1.55)
    (0.98,1.8)
    (0.99,2.27)
    (1,2.67)
}; 
\end{axis}
\node (a1) at (.5, .8\linewidth){\bf (a) QENP};
\node (a1) at (.5, .5\linewidth){\bf (b) MFNP};
\node (a1) at (.5, .2\linewidth){\bf (c) SWS};
\begin{axis}[
  at={(0.5\linewidth, .6\linewidth)},
  ybar=0,
  bar width=5,
  ylabel style={align=center},
  legend style={
    at={(1.1,1.6)},
    legend columns=1,
    font=\small,draw=none,fill=none},
  xmax=30,xmin=1,
]
\addplot+[black, fill=gray!60, draw=black, thick,area legend]
  coordinates{
    (5,1.11)
    (10,1.17)
    (15,1.17)
    (20,1.12)
    (25,1.16)
}; \addlegendentry{Average};
\addplot+[black, fill=gray, draw=black, thick,area legend]
  coordinates{
    (5,1.71)
    (10,1.73)
    (15,2)
    (20,2.67)
    (25,2.67)
}; \addlegendentry{Max};
\end{axis}
\begin{axis}[
  at={(0.5\linewidth, 0.3\linewidth)},
  ybar=0,
  bar width=5,
  legend style={at={(0,1.4)},legend columns=2,font=\small,draw=none,fill=none},xmax=30,xmin=1,
]
\addplot+[black, fill=gray!60, draw=black, thick,area legend]
  coordinates{
    (5,1.11)
    (10,1.17)
    (15,1.17)
    (20,1.12)
    (25,1.16)
};
\addplot+[black, fill=gray, draw=black, thick,area legend]
  coordinates{
    (5,1.71)
    (10,1.73)
    (15,2)
    (20,2.67)
    (25,2.67)
}; 
\end{axis}
\begin{axis}[
  at={(0.5\linewidth, 0)},
  ybar=0,
  xlabel=PWL Segments, 
  bar width=5,
  legend style={at={(1.1,1.3)},legend  columns=1,font=\small,draw=none,fill=none},
  xmax=30,xmin=1,
]
\addplot+[black, fill=gray!60, draw=black, thick, area legend]
  coordinates{
    (5,1.07)
    (10,1.09)
    (15,1.095)
    (20,1.09)
    (25,1.1)
}; 
\addplot+[black, fill=gray, draw=black, thick, area legend]
  coordinates{
    (5,1.7)
    (10,1.78)
    (15,1.6)
    (20,1.6)
    (25,1.6)
}; 
\end{axis}
\node (a1) at (.55\linewidth, .8\linewidth){\bf (d) QENP};
\node (a1) at (.55\linewidth, .5\linewidth){\bf (e) MFNP};
\node (a1) at (.55\linewidth, .2\linewidth){\bf (f) SWS};
\end{tikzpicture}

%% file: figures/pwl_functions.tikz
\begin{tikzpicture}
\pgfplotsset{
  width=0.55\linewidth,
  height=0.35\linewidth,
  xtick pos=left,
  ytick pos=left,
  ymajorgrids=true,
  tick label style={font=\small},
  label style={font=\small},
  xmax=26,xmin=4,
}
\begin{axis}[
  at={(0\linewidth, 0)}, 
  xlabel=Segments, 
  ylabel={Runtime (s)},
  bar width=5,
  legend style={
    at={(1.9,1.35)},
    legend columns=3,
    font=\small,draw=none,fill=none},
]
\addplot+[mark=triangle*,thick]
  coordinates{
    (5, 0.1)
    (10, 0.4)
    (15, 0.65)
    (20, 0.95)
    (25, 2)
};\addlegendentry{MFNP};
\addplot+[thick]
  coordinates{
    (5, 0.18)
    (10, 0.44)
    (15, 0.53)
    (20, 0.85)
    (25, 0.9)
};\addlegendentry{QENP};
\addplot+[thick]
  coordinates{
    (5, 0.5)
    (10, 3.5)
    (15, 8)
    (20, 11)
    (25, 13)
};\addlegendentry{SWS};
\end{axis}
\node (a1) at (.2, -.6){\bf (a)};
\begin{axis}[ 
  at={(0.5\linewidth, 0)},
  xlabel=Segments,
  ylabel=Utility,
  legend style={
    at={(1.1,1.3)},
    legend columns=3,
    font=\small,
    draw=none,fill=none
  },
  ymin=0
]
\addplot+[mark=triangle*,thick]
  coordinates{
    (5, 3.5)
    (10, 4.01)
    (15, 4.4)
    (20, 4.55)
    (25, 4.6)
};
\addplot+[thick]
  coordinates{
    (5, 2.6)
    (10, 3.6)
    (15, 3.7)
    (20, 3.9)
    (25, 3.9)
};
\addplot+[thick]
  coordinates{
    (5, 0.35)
    (10, 0.4)
    (15, 0.44)
    (20, 0.45)
    (25, 0.46)
};
\end{axis}
 \node (a1) at (.55\linewidth, -.6){\bf (b)};
\end{tikzpicture}

%% file: 6-FieldTests.tex
\section{Field Tests}
Previous studies evaluated predictive modeling for anti-poaching with field tests in QENP~\cite{kar2017cloudy, gholami2018adversary}. However, before scaling PAWS globally to over 800 parks worldwide through integration with SMART, it is critical to test these algorithms on the ground in other protected areas to ensure strong performance across diverse environments. To evaluate the predictive component of our model, we present results from deploying PAWS in two protected areas: MFNP in Uganda and SWS in Cambodia. 

\subsection{Field Tests in Murchison Falls National Park}
In collaboration with the Uganda Wildlife Authority, we conducted field tests in Murchison Falls National Park to compare with previous results from Queen Elizabeth. QENP and MFNP are both in Uganda but 500~km apart. They are both large savannas, but QENP contains more scrub and woodland while MFNP has large grasslands. The shape of QENP is long so it is easy to access the center from the boundary, while MFNP is circular with a more protected core so most poaching occurs at the edges of the park. From a computational perspective, the class imbalance differs between the parks, with MFNP having a threefold higher proportion of positive labels~(see Table~\ref{table:datasets}).



To conduct these field tests, we selected $2 \times 2$~km regions and classified our recommended areas into three risk groups (low, medium, and high). These areas were all infrequently patrolled in the past, to ensure we test the predictive power of our algorithms rather than relying on past patterns, as \cite{kar2017cloudy} did. To reduce bias in data collection, we did not reveal the risk groups to rangers before conducting the experiments. The predictive scores for this field test were selected using the iWare-E model with bagging decision trees. When selecting risk group, we used the prediction of the model at a nominal patrol effort, which the rangers will be likely able to achieve. 

\begin{figure}
 \input{figures/field_tests.tikz}
 \caption{Detected instances of poaching per square kilometer patrolled across high-, medium-, and low-risk regions in the four field tests. (\# Obs. / \# Cells in Table~\ref{table:fieldTestResults})}
 \label{fig:fieldTestGraph}
\end{figure}
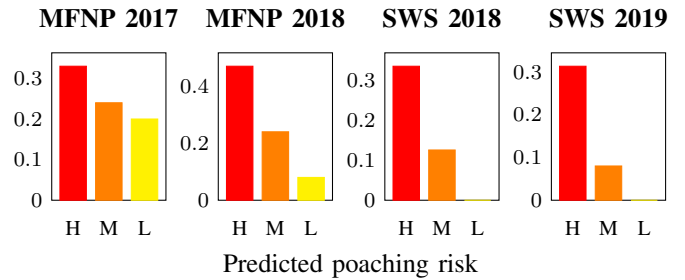

Table~\ref{table:fieldTestResults} summarizes the field test results for MFNP, with the results \# Obs. / \# Cells visualized in Fig.~\ref{fig:fieldTestGraph} for increased clarity. These experiments were executed for 5~months, during which park rangers detected poaching activity in 38~cells. Since the predictive models are trained based on data with three-month time resolution, we present the field test results as two- and three-month groups. We report the number of cells in which poaching activity was observed (\#~Obs.), number of $1 \times 1$~km cells patrolled (\#~Cells), and amount of patrol effort in km (Effort). Due to limited park ranger resources, not all the selected blocks were patrolled. We thus count the number of cells that were actually patrolled to compute the normalized number of observations (\#~Obs.~/~\#~Cells). We normalize by number of cells rather than unit of effort because our model makes binary predictions of poaching risk in each cell, not as a function of effort.

Although we did not inform park rangers of the risk classification of each block when conducting the field tests, the results indicate that park rangers expended more effort in high-risk areas than low-risk ones. This suggests that they had an intuitive understanding of which regions were more worthwhile to patrol and thus allocated more effort there, indicating the our predictive models aligns with the expert knowledge of the domain experts. 

The average incidences of poaching activity observed is highest in the high-risk regions and lowest in the low-risk regions for both time steps, indicating that our algorithm effectively learns which areas are more at risk of poaching. We use a Pearson's chi-squared test to assess independence of the observations on two variables (\# Obs. and Risk group in Table~\ref{table:fieldTestResults}). The chi-squared test on all the experimental results yields a $p$-value of $1.05 \times 10^{-2}$, which is statistically significant at a level of 0.05.



\begin{table}
\centering
\caption{Field test results}
\label{table:fieldTestResults}
\begin{tabular}{c|cccc}
\Xhline{2\arrayrulewidth}
 \textbf{Risk group} &\textbf{\# Obs.} & \textbf{\# Cells}  & \textbf{Effort}  & \textbf{\# Obs. / \# Cells} \\
\hline
& \multicolumn{4}{c}{\textbf{MFNP trial 1: Nov--Dec 2017}} \\ \hline
  \textbf{High}   &6	&18 &71.6	&0.33 \\
  \textbf{Medium} &5	&21 &31.9  &0.24 \\
  \textbf{Low}    &2	&10 &12.6  &0.20 \\
\Xhline{2\arrayrulewidth}
& \multicolumn{4}{c}{\textbf{MFNP trial 2: Jan--Mar 2018}} \\ \hline
  \textbf{High}   &17 &36 &197.4 &0.47  \\
  \textbf{Medium} &7  &34 &83.4 &0.21  \\
  \textbf{Low}    &1  &13 &45.1  &0.08  \\
\Xhline{2\arrayrulewidth}
\multicolumn{5}{c}{} \\
\Xhline{2\arrayrulewidth}
& \multicolumn{4}{c}{\textbf{SWS trial 1: Dec 2018--Jan 2019}}
\\ \hline
\textbf{High}     &14	& 42 &103.8  &0.34 \\
\textbf{Medium}   &5	& 40 &111.36  &0.13 \\
\textbf{Low}      &0	& 36 &84.12 &0 \\
\Xhline{2\arrayrulewidth}
& \multicolumn{4}{c}{\textbf{SWS trial 2: Feb--Mar 2019}}
\\ \hline
\textbf{High}     &7	& 22 &71.52  &0.31 \\
\textbf{Medium}   &2	& 25 &69.58  &0.08 \\
\textbf{Low}      &0	& 34 &89.07 &0 \\
\Xhline{2\arrayrulewidth}
\end{tabular}
\end{table}

\subsection{Field Tests in Srepok Wildlife Sanctuary}
\label{sec:sws_field_test}

We deployed the enhanced iWare-E predictive algorithm in Southeast Asia for the first time. Partnering with WWF Cambodia, we conducted field tests in Srepok Wildlife Sanctuary beginning in December 2018. As discussed in Section~\ref{sec:cambodia_challenges}, this park differs in many ways from MFNP and QENP such as significantly greater class imbalance (only 3.6\% positive labels), denser ecology and terrain, and more limited patrolling resources. 

\begin{figure}
\centering
\includegraphics[width=.8\linewidth]{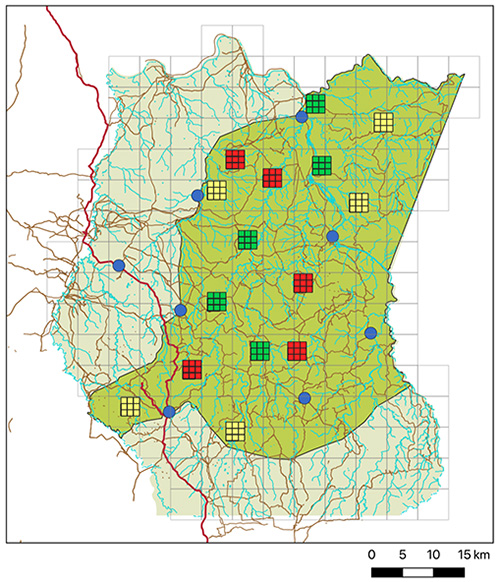}
\caption{Regions in SWS used for field tests in December 2018. High-, medium-, and low-risk regions are shown in red, yellow, and green, respectively. Blue circles are patrol posts; the rivers and roads are also displayed.}
\label{fig:SWS_test_areas}
\end{figure}

Since SWS experiences high seasonality, where many rivers dry up during the dry season, we trained our model based only on data from dry months (November through April), using a two-month discretization to get three temporal points per year. We built an iWare-E model, using the enhanced approach described in this paper, with GPs as the weak learner trained on data from January 2015 through April 2018. We made predictions on the November--December 2018 time period for the December--January field tests, and then predicted on January--February 2019 for the February--March field tests. Using these predictions of poaching risk for $1 \times 1$~km cells, we averaged the risk predictions over the adjacent cells by convolving the risk map to produce $3 \times 3$~km blocks. We then discarded all blocks with historical patrol effort above the 50th percentile, to ensure we were assessing the ability of our model to make predictions in regions with limited data. From this set of valid blocks, we identified high-, medium-, and low-risk areas by considering blocks with risk predictions within the 80--100, 40--60, and 0--20 percentile. We selected the final areas used during the field tests based on criteria the park manager specified, including proximity to roads and being within a high-protection core zone. In total, we selected five $3 \times 3$~km blocks from each of the three risk categories, shown in Fig.~\ref{fig:SWS_test_areas}. To deploy these tests, we gave the park rangers GPS coordinates of the center of each block and asked them to target those regions during their patrols. Again, we did not reveal to them the risk category of each region to prevent bias. 

In December~2018, 72~park rangers in teams of eight began conducting patrols throughout the park, focusing on our suggested areas. The results from this real-world experiment are presented in Table~\ref{table:fieldTestResults}. 
A chi-squared test confirms that the results are statistically significant at the 0.05 level with $p$-values of $2.3 \times 10^{-2}$ and $0.7 \times 10^{-2}$ respectively. These results bolster our findings from the MFNP experiments: our predictive model effectively evaluates the poaching threat for different regions across the park, to even greater success at discriminating between high-risk and low-risk areas. \textit{Park rangers found absolutely no poaching activity in low-risk areas, despite exerting a comparable amount of effort in those regions.} This result suggests that revealing our risk predictions to them would grant them valuable insight into their patrol strategy so as to more effectively allocate their limited resources, as they can confidently spend less time patrolling these low-risk regions.

During the first month of field tests in SWS, park rangers detected poaching in 32~cells and removed over one thousand snares---each of which represents the life of one animal potentially saved. The rangers informed us that, by comparison, they typically remove two or three hundred snares in an average month. In that month, they also confiscated a firearm and saw, but did not catch, three groups of poachers in the forest. We are enthusiastic about these results, and believe that a data-driven approach can bolster wildlife protection efforts. 

\subsection{Lessons Learned from Deployment}

In the data-to-deployment pipeline, it is critical to work with collaborators throughout the process. 

Extended discussions with park rangers made us aware of how much patrols may change over time. First, changes in funding and support may drastically alter patrolling resources; the number of park rangers in SWS in 2018 is over double that of 2016. Second, park rangers must be sensitive to real-time changes in the illegal market. For example, illegal logging rampantly increased in a nearby park in March, so park rangers from SWS were redirected there to provide backup, reducing the sources inside the park. These examples emphasize the importance of conducting self-contained experiments when evaluating model performance. As mentioned, the first field test in QENP~\cite{kar2017cloudy} suggested areas to park rangers for one month of patrols, then compared the number of findings to historical averages. However, the increased findings may be partially explained by factors beyond the predictive model. 

The insight that poaching activity is subject to seasonality also resulted from discussions with park rangers, which inspired us to break up the SWS data into rainy vs. dry season and generate those predictive models separately. Our predictive model identified higher poaching risk in the north during dry season and south during rainy season, which garnered immediate positive feedback from the rangers: this aligned with their experiences on the ground and understanding of the accessibility of the terrain.


Finally, real-world deployment is necessary for effective technology transfer. Park rangers are motivated by positive conservation outcomes, not by improvements in AUC. We are pleased to report that after two months in SWS, our partners at WWF were thrilled by the results and eager to deploy in other parks around the world. These field tests in SWS are ongoing, and we will continue to improve the PAWS system based on our findings and discussions with domain experts.


%% file: figures/field_tests.tikz
\begin{tikzpicture}
\pgfplotsset{
    width=0.35\linewidth,
    height=0.4\linewidth,
    xtick={1,2,3},
    xticklabels={H, M, L},
    xtick pos=left,
    ytick pos=left,
    ybar,
    ymin=0,
    tick label style={font=\footnotesize},
    x tick style={draw=none},
    y tick style={draw=none},
    enlarge x limits=.3,
    every axis plot/.append style={
        ybar,
        bar width=10,
        bar shift=0pt,
        fill,
    }
}
\begin{axis}[
    at={(0\linewidth, 0)},
    title={\textbf{MFNP 2017}},
]
\addplot[red]coordinates{(1,0.33)};
\addplot[orange]coordinates{(2,0.24)};
\addplot[yellow]coordinates{(3,0.20)};
\end{axis}
\begin{axis}[
    at={(.25\linewidth, 0)},
    title={\textbf{MFNP 2018}},
]
\addplot[red]coordinates{(1,0.47)};
\addplot[orange]coordinates{(2,0.24)};
\addplot[yellow]coordinates{(3,0.08)};
\end{axis}
\begin{axis}[
    at={(.5\linewidth, 0)},
    title={\textbf{SWS 2018}},
    xlabel={Predicted poaching risk},
    xlabel style={at={(axis description cs:-.3,-.3)}},
]
\addplot[red]coordinates{(1,0.336)};
\addplot[orange]coordinates{(2,0.126)};
\addplot[yellow]coordinates{(3,0.0)};
\end{axis}
\begin{axis}[
    at={(.75\linewidth, 0)},
    title={\textbf{SWS 2019}},
]
\addplot[red]coordinates{(1,0.313)};
\addplot[orange]coordinates{(2,0.080)};
\addplot[yellow]coordinates{(3,0.0)};
\end{axis}
\end{tikzpicture}

%% file: 8-Conclusion.tex
\section{Conclusion}


To assist park rangers in combating illegal wildlife poaching, we take an end-to-end approach to the data-to-deployment pipeline by identifying an unaddressed need: to conduct risk-averse patrols and maximize the number of snares removed, a challenge exacerbated by uncertainty and bias in their historical data. Thus, we enhance PAWS by accounting for predictive uncertainty when computing patrol paths. To do so, we integrate Gaussian processes into an enhanced iWare-E model to quantify uncertainty in the predictions of poaching risk. We then exploit this uncertainty in the patrol planning model to improve robustness of the prescribed plans, increasing detection of poaching by an average of 30\%. Additionally, we present results from extensive field tests in Uganda and Cambodia, demonstrating that our predictive model strongly discriminates between relative threat of poaching and can be effectively applied on the ground. During these tests, rangers detected 38 attacked cells in MFNP and confiscated over 1,000 snares in SWS in a single month. These real-world successes indicate that a data-to-deployment pipeline can be effectively implemented for wildlife protection, and ought to be extended to other domains.

\section*{Acknowledgment}
Thank you to our partners at Uganda Wildlife Authority, Cambodia Ministry of Environment, Wildlife Conservation Society, and the World Wide Fund for Nature for their conservation efforts. We are grateful to the rangers in MFNP and SWS for supporting our field tests, especially in collecting and providing data from their patrols. Special thanks to Alexander Wyatt, James Lourens, and WWF Cambodia for their work in SWS. 

This work was supported by the Army Research Office (MURI W911NF1810208).